\documentclass[aps,prl,twocolumn,superscriptaddress,longbibliography,nobalancelastpage]{revtex4-2}

\usepackage{lipsum}
\usepackage{graphicx}
\usepackage{amsmath}
\usepackage{amssymb}
\usepackage{color}
\usepackage{enumitem}
\usepackage[normalem]{ulem}
\usepackage[utf8]{inputenc}
\usepackage{environ}
\usepackage[dvipsnames]{xcolor}
\usepackage[colorlinks = true,allcolors = {Blue}]{hyperref}
\usepackage{orcidlink}

\begin{document}
\title{Entanglement--magic separation in hybrid quantum circuits}
\author{Gerald E. Fux\,\orcidlink{0000-0002-7912-0501}}
\affiliation{The Abdus Salam International Center for Theoretical Physics (ICTP), Strada Costiera 11, 34151 Trieste, Italy}
\author{Emanuele Tirrito\,\orcidlink{0000-0001-7067-1203}}
\affiliation{The Abdus Salam International Center for Theoretical Physics (ICTP), Strada Costiera 11, 34151 Trieste, Italy}
\affiliation{Pitaevskii BEC Center, CNR-INO and Dipartimento di Fisica,
Università di Trento, Via Sommarive 14, Trento, I-38123, Italy}
\author{Marcello Dalmonte\,\orcidlink{0000-0001-5338-4181}}
\affiliation{The Abdus Salam International Center for Theoretical Physics (ICTP), Strada Costiera 11, 34151 Trieste, Italy}
\affiliation{Scuola Internazionale Superiore di Studi Avanzati (SISSA), Via Bonomea 265, 34136 Trieste, Italy}
\author{Rosario Fazio\,\orcidlink{0000-0002-7793-179X}}
\affiliation{The Abdus Salam International Center for Theoretical Physics (ICTP), Strada Costiera 11, 34151 Trieste, Italy}
\affiliation{Dipartimento di Fisica ``E. Pancini", Universit\`a di Napoli ``Federico II'', Monte S. Angelo, I-80126 Napoli, Italy}
\date{\today}

\begin{abstract}
    Magic describes the distance of a quantum state to its closest stabilizer state.
    It is---like entanglement---a necessary resource for a potential quantum advantage over classical computing.
    We study magic, quantified by stabilizer entropy, in a hybrid quantum circuit with projective measurements and a controlled injection of non--Clifford resources.
    We discover a phase transition between a (sub)--extensive and area law scaling of magic controlled by the rate of measurements.
    The same circuit also exhibits a phase transition in entanglement that appears, however, at a different critical measurement rate.
    This mechanism shows how, from the viewpoint of a potential quantum advantage, hybrid circuits can host multiple distinct transitions where not only entanglement, but also other non--linear properties of the density matrix come into play.
\end{abstract}

\maketitle


The resources of quantum physics allow to achieve an advantage over classical computing.
It is well known that entanglement is a necessary resource to achieve that goal, but it is, however, not sufficient.
Clifford circuits may lead to highly entangled states but can nonetheless be simulated classically in polynomial time~\cite{gottesman_heisenberg_1998, aaronson_improved_2004, veitch_negative_2012}.
They are generated by the non--universal gate set of the Hadamard, $\pi/4$ phase, and controlled--not gates.
The states that can be reached with Clifford circuits starting from states in the computational basis are called \emph{stabilizer states} and they play an important role in the field of error correction as they quite naturally lead to a large and powerful class of error correcting codes~\cite{nielsen_quantum_2010, gottesman_class_1996, gottesman_stabilizer_1997, gottesman_theory_1998}.
Adding any non--Clifford gate, such as the T--gate ($\pi/8$ phase gate) to the Clifford gate set makes it universal and thus allows to reach any state.
Because the stabilizer states can be efficiently simulated with a classical computer despite their potentially highly entangled nature, any quantum circuit that should potentially achieve quantum advantage must therefor not only include highly entangled states but also include non--Clifford gates.
The amount of non--Clifford resources necessary to create a state is called \emph{non--stabilizerness}, or \emph{magic}~\cite{bravyi_universal_2005, veitch_resource_2014, chitambar_quantum_2019, liu_many-body_2022}.
Furthermore, a deepened understanding of magic is particularly relevant in the development of fault--tolerant universal quantum computing, where Clifford gates may be implemented straight forwardly in a fault--tolerant way, but non--Clifford gates are more challenging and require magic state distillation or similar procedures~\cite{campbell_roads_2017}.

\begin{figure}[t]
	\includegraphics[width=0.49\textwidth]{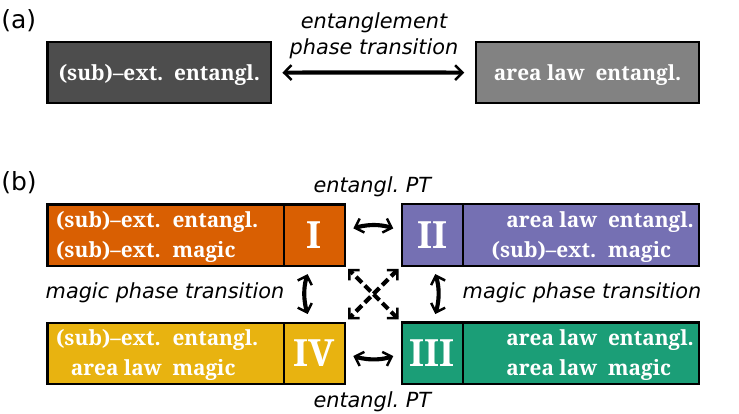}
	\caption{\label{fig:magic-and-entanglement}%
		Phase transitions in entanglement~(a) compared to the possible phase transitions in entanglement \emph{and} magic~(b) in one--dimensional quantum circuits.
		Quantum circuits in phase~I may lead to an advantage over classical computing, while circuits in phases~II and~III are amenable to simulation through tensor network methods, and circuits in phases~III and~IV can be tackled efficiently with the stabilizer formalism.
	}
\end{figure}%

The importance of entanglement has stimulated intense interest in its role not only in quantum information but also in other fields reaching from condensed matter to statistical mechanics and the quantum mechanics of black holes~\cite{calabrese_entanglement_2004, popescu_entanglement_2006, hayden_black_2007, amico_entanglement_2008, horodecki_quantum_2009, laflorencie_quantum_2016, almheiri_entropy_2019}.
In some many body systems entanglement may display critical behavior even when all state--observables are featureless.
One example of this are hybrid quantum circuits that consist of unitary gates interspersed with measurements~\cite{potter_entanglement_2022, skinner_measurement-induced_2019, li_quantum_2018, li_measurement-driven_2019}.
These circuits can exhibit so--called Measurement Induced Phase Transitions (MIPTs) in entanglement.
There are multiple different circuits which show such MIPTs in entanglement, such as random Clifford~\cite{li_quantum_2018,li_measurement-driven_2019} or Haar--random circuits~\cite{skinner_measurement-induced_2019} interspersed with projective measurements.
Both of these examples show the same qualitative behavior of entanglement, namely a volume law below a critical measurement rate and an area law above~\cite{sierant_universal_2022}.
However, from the perspective of a potential quantum advantage these two examples are very different.
This is because even in the entanglement volume law phase the random Clifford circuit can be efficiently simulated classically using the stabilizer formalism, while the same is not true for the Haar--random circuit.
Furthermore, MIPTs in entanglement have been related to the performance of quantum error correction codes, for which it has been found that hybrid circuits that show a volume law in entanglement encode quantum information, while circuits that show an area law destroy it irreversibly~\cite{gullans_dynamical_2020}.
In this context, magic could further differentiate between quantum error correction codes that require non--Clifford resources (which are hard to implement in a fault--tolerant way) and codes that do not.
It is thus natural to ask whether hybrid quantum circuits can also show MIPTs in magic, and if yes, how those phase transitions relate to the known MIPTs in entanglement.

In this Letter we present a new kind of phase transition between a (sub)--extensive and area law scaling of magic in a hybrid quantum circuit controlled by the rate of measurement.
Most notably, we find that this phase transition appears at a different critical measurement rate compared to the critical rate for the volume to area law transition of entanglement.
This suggest that the mechanism that drives the observed magic phase transition is different from the mechanism driving the entanglement phase transition.
From the perspective of a potential quantum advantage it is thus necessary to extend the picture of MIPTs in entanglement (see Fig.~\ref{fig:magic-and-entanglement}a) to MIPTs in entanglement \emph{and} magic (see Fig.~\ref{fig:magic-and-entanglement}b).
In particular, the hybrid quantum circuit studied in this Letter shows an entanglement phase transition from a (sub)--extensive to area law while the magic remains (sub)--extensive in both phases [phase I $\rightarrow$ phase II], as well as a magic phase transition from a (sub)--extensive to area law while the entanglement keeps an area law behavior [phase II $\rightarrow$ phase III].
We note that in this paper we use the term ``area law'' for convenience to denote a constant scaling with system size, and that for magic there is no expectation of an according scaling in quantum circuits of higher dimensions.

While phase transitions in magic have been reported previously in Refs.~\cite{leone_phase_2023}~and~\cite{niroula_phase_2023}
we argue that the setup and the consequent phase transitions observed in this work are qualitatively different.
The work by Leone \textit{et al.}~\cite{leone_phase_2023} studies to what extend Clifford circuits interspersed with T--gates can be transformed into circuits with all T--gates only in a small subset of qubits and does not involve any measurements.
In Ref.~\cite{niroula_phase_2023} Niroula \textit{et al.} study a random Clifford error correcting protocol with a stabilizer state as an input and a noise layer that injects magic through phase gates at a variable angle.
This setup shows a phase transition of magic in the output state depending on the phase gate angle.
Further analysis reported by Turkeshi and Sierant~\cite{turkeshi_error-resilience_2023} suggests that this phase transition stems from a phase transition in the success of error correction.
Following from this, the transition presented in Ref.~\cite{niroula_phase_2023} could thus be witnessed through many different properties of a state, including magic, but also any other property, such as entanglement, would exhibit the same critical measurement rate set by the error correcting transition.
Hence, to the best of our knowledge our work is the first to explicitly show the existence of an MIPT in magic that is independent of an MIPT in entanglement.


\begin{figure}[t]
	\includegraphics[width=0.49\textwidth]{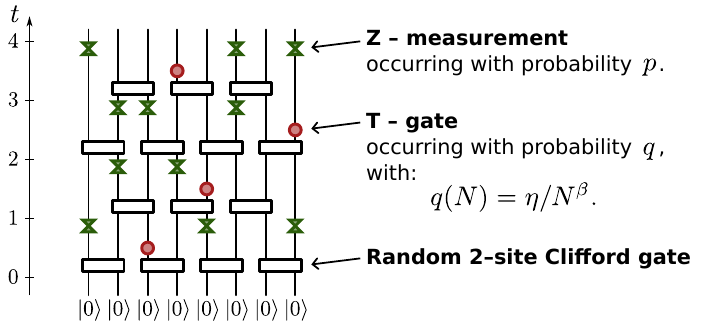}
	\caption{\label{fig:circuit}%
		Hybrid quantum circuit with a brickwork structure of random 2--site Clifford gates interspersed with measurements in the computational bases and T--gates for $N=8$ qubits and 4 time steps. The measurements and T--gates appear randomly with probability $p$ and $q(N)=\eta/N^\beta$, respectively, where $\beta$ determines the scaling of the T--gate density with prefactor $\eta$.
	}
\end{figure}%

\textit{The setup.}---%
To find a phase transition in magic we study hybrid quantum circuits that interpolate between situations found in random Clifford and Haar--random hybrid circuits in a controlled way.
For this we intersperse a brickwork of random two--site Clifford gates with T--gates in addition to the projective measurements.
Figure~\ref{fig:circuit} shows an example realization of such a circuit for 8 qubits and 4 time steps.
The measurements occur on any of the $N$ qubits at any time step with a uniform probability $p$, while the T--gates occur with a uniform probability $q$ which we set to scale as $q(N) = \eta / N^\beta$.
Beside the prefactor $\eta$, also the exact exponent $\beta$ of the scaling will play an important role, as discussed further below.
For any realization of such a circuit there is an ensemble of pure quantum state trajectories, where each trajectory is labeled by the measurement outcome sequence.
We are interested in the long time limit of entanglement and magic averaged over both the quantum circuit realizations and the measurement outcomes.
Both entanglement and magic are non--linear in the density matrix and it is thus important to distinguish between the entanglement/magic of the average state versus the average of the entanglement/magic over the ensemble of trajectories.
While the former is not even well defined (given that we use measures of entanglement/magic applicable only to pure states), the latter may exhibit interesting behavior.


\textit{Entanglement and stabilizer entropy.}---%
As a measure for entanglement $\mathcal{E}$ we choose the von--Neumann entropy between the left and right half of the chain.
Given an $N$ qubit pure state $|\psi\rangle$ it is
\begin{equation}
\mathcal{E}(|\psi\rangle) = - \mathrm{tr}\left[ \rho_L \log_2(\rho_L) \right]\mathrm{,}
\end{equation}
where $\rho_L = \mathrm{tr}_R\left[|\psi\rangle\langle\psi|\right]$ is the reduced density matrix of the left half of the chain $L=\{1, \ldots, N/2-1\}$ with $\mathrm{tr}_R$ denoting the partial trace over the right half $R = \{N/2, \ldots, N\}$.
While it is generally undisputed that the von--Neumann entropy is a good measure for bipartite entanglement in pure states, the choice of a ``good measure'' $\mathcal{M}$ for magic is under discussion in current literature~\cite{%
veitch_resource_2014,
howard_application_2017,
heinrich_robustness_2019,
wang_quantifying_2019,
bu_efficient_2019,
beverland_lower_2020,
sarkar_characterization_2020,
jiang_lower_2023,
leone_stabilizer_2022,
oliviero_measuring_2022,
bu_complexity_2022,
haug_stabilizer_2023,
bu_quantum_2023,
tirrito_quantifying_2023,
turkeshi_measuring_2023,
bu_stabilizer_2023,
gu_little_2023}.
In this work we choose the recently introduced stabilizer entropy as a measure of magic~\cite{leone_stabilizer_2022}.

The stabilizer entropy quantifies the spread of a state in the basis of Pauli string operators.
In particular, the stabilizer $\alpha$-R\'{e}nyi entropy of an $N$ qubit pure state $|\psi\rangle$ is
\begin{equation} \label{eq:sre-def}
\mathcal{M}(|\psi\rangle) = \frac{1}{1-\alpha} \log_2 \left( \sum_{P\in \mathcal{P}_N} \Xi_P^\alpha(|\psi\rangle) \right) - N \mathrm{,}
\end{equation}
where $\Xi_P(|\psi\rangle) = 2^{-N} \langle\psi| P |\psi\rangle^2$ is the probability distribution over all Pauli strings  $P \in \mathcal{P}_N$ with +1 phases.
This definition is motivated by the fact that a state is a stabilizer state if and only if there are $2^N$ Pauli strings $P\in \mathcal{P}_N$ with an expectation value of 1 or -1, while the expectation for all others is 0.
Below we present numerical results for $\alpha=2$, but note that we find the same critical measurement rates for the limit of $\alpha\rightarrow 1$ (which yields the stabilizer Shannon entropy).
One advantage of the stabilizer entropy over many other proposed measures of magic is that it allows an efficient computation even for a large number of qubits through Monte Carlo sampling of the Pauli string probability distribution~\cite{haug_quantifying_2023, lami_quantum_2023, tarabunga_many-body_2023}.


\begin{figure}[t]
	\includegraphics[width=0.45\textwidth]{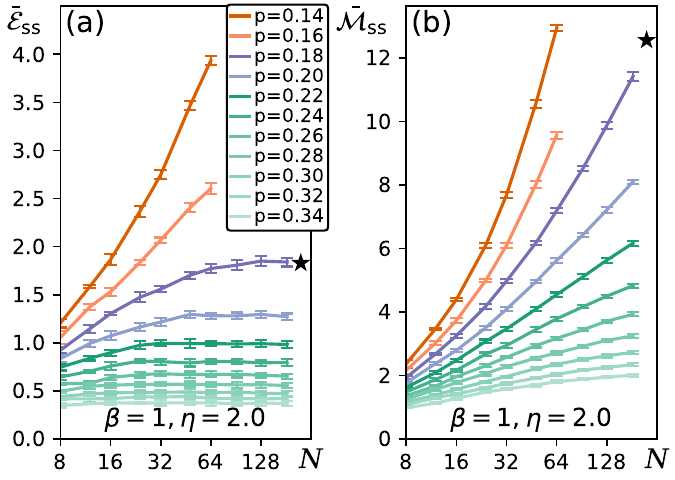}
	\caption{\label{fig:entropies-vs-size}%
		Numerical results for the average steady state entanglement~(a) and magic~(b) versus system size $N$ for several different measurement rates $p$ and T--gate density $q(N) = \eta/N^\beta$ with $\eta=2.0$ and $\beta=1$.
		Note that for the measurement rate $p=0.18$ (marked with $\bigstar$) the lines bend downwards in panel~(a) but upwards in panel~(b), which indicates an area law in entanglement but a (sub)--extensive law in magic.
	}
\end{figure}%

\textit{Numerical results for the $\beta=1$ case.}---%
As mentioned above, the scaling of the T--gate density with system size plays an important role for a potential phase transition in magic.
In particular, for a constant density of T--gates ($\beta=0$) one expects a trivial volume law scaling of magic for any measurement rate $p<1$.
This is because magic is a global property, but---unlike entanglement---it can be created locally.
Because of the additivity of magic~\cite{leone_stabilizer_2022} any finite magic density will thus lead to a volume law in magic for the total state.
In contrast to this, entanglement occurs between at least two subsystems and may only be created through gates that lie on the boundary of those subsystems.
In the setup studied here there is only one two--site Clifford gate every other time step that can increase the entanglement between the left and right half of the chain, independently of the system size.
In analogy, this naturally suggests to study magic in a quantum circuit with a constant average number $\eta$ of  T--gates per time step, also independently of the system size.
This corresponds to a vanishing T--gate density $q=\eta/N^\beta$ with $\beta=1$.
In the following we thus start by studying the case of $\beta=1$, but we will also briefly discuss other scalings further below.

\begin{figure}[t]
	\includegraphics[width=0.35\textwidth]{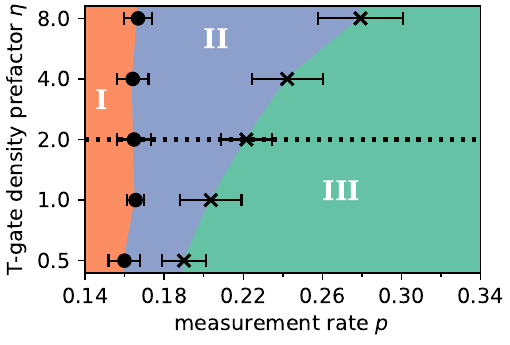}
	\caption{\label{fig:phase-diagram}%
		Phase diagram ($\eta$ vs. $p$) extracted from the numerical results for $\beta=1$ and $\eta=\{0.5,\ldots,8.0\}$.
		The bullets and crosses show the position of the entanglement and magic phase transition, respectively.
            We present the details of the numerical error estimation (displayed as error bars) in the SM~\cite{sm_note}.
            The colored regions resulting from this data correspond to the phases I, II, and III shown in Fig.~\ref{fig:magic-and-entanglement}.
		For low measurement rates we find that both entanglement and magic follow a (sub)--extensive scaling (phase I), while for high measurement rates both follow an area law (phase III).
		However, we also find an extended region in between in which entanglement scales as an area law, but magic scales (sub)--extensively (phase II).
		The dotted line corresponds to the data shown in Fig.~\ref{fig:entropies-vs-size}.
	}
\end{figure}%

Figure~\ref{fig:entropies-vs-size} shows the growth of the trajectory averaged entanglement and magic with system size (up to 184 sites) for $\beta=1$, $\eta = 2.0$, and several different measurement rates $p$.
To generate this graph we simulate 512 trajectories using time evolving block decimation~\cite{vidal_efficient_2004, fishman_itensor_2022} for each parameter set of $p$ and $N$.
For each of those trajectories we periodically (every 8$^\mathrm{th}$ time step) evaluate the von--Neumann entropy $\mathcal{E}(t)$ and the stabilizer 2--R\'enyi entropy $\mathcal{M}(t)$ using the method introduced in~\cite{lami_quantum_2023}.
Then we yield $\bar{\mathcal{E}}(t)$ and $\bar{\mathcal{M}}(t)$ from the average of those values.
For the parameter sets considered here we find that $\bar{\mathcal{E}}(t)$ and $\bar{\mathcal{M}}(t)$ attain an approximate steady value within less than $2N$ time steps.
To get a more accurate approximation of the steady values $\bar{\mathcal{E}}_\mathrm{ss}$ and $\bar{\mathcal{M}}_\mathrm{ss}$, we compute the dynamics for $4N$ time steps and average over $\bar{\mathcal{E}}(t)$ and $\bar{\mathcal{M}}(t)$ for all time steps between $2N$ and $4N$.
Finally we use the spread of this distribution as an estimate for the error due to the finite number of sampled trajectories, which we display as error bars in Fig.~\ref{fig:entropies-vs-size}.
We present further details on the numerical computation and error estimation in the supplemental material~(SM)~\cite{sm_note}.

The data shown in Fig.~\ref{fig:entropies-vs-size} strongly suggests the existence of \emph{two} distinct MIPTs.
Figure~\ref{fig:entropies-vs-size}a shows that in this log--linear plot for $\bar{\mathcal{E}}_\mathrm{ss}(N)$, the line for $p=0.14$ tends upwards, while the line for $p=0.16$ appears approximately straight, and the lines for $p\geq 0.18$ bend downwards. This indicates a (sub)--extensive/log/area law in entanglement, respectively.
This observation is consistent with the known measurement induced phase transition of entanglement for Clifford circuits, for which the addition of a vanishing density of T--gates appears not to significantly modify the critical measurement rate \mbox{$p_c^\mathrm{entgl.}=0.15995(10)$}~\cite{li_quantum_2018, sierant_measurement-induced_2022}.
On the other hand, the values of magic for the same trajectories plotted in Fig.~\ref{fig:entropies-vs-size}b also indicate (sub)--extensive/log/area law, but with a significantly higher critical measurement rate $p_c^\mathrm{magic} \simeq 0.22$.
In particular, we note that for $p=0.18$ (marked with $\bigstar$) the results clearly indicate an area law for entanglement but a (sub)--extensive law for magic.

As the scaling of magic near the critical point is yet unknown we do not perform any scaling collapse in this work.
We instead focus on the extraction of the critical measurement rates from the data shown in Fig.~\ref{fig:entropies-vs-size} by fitting each curve with two different functions,  corresponding to a (sub)--extensive and area law respectively and perform a so--called F--test~\cite{hahs-vaughn_introduction_2020}.
For this, we fit $a$, $b$, and $\gamma$ in the function $f(N) = a + b N^\gamma$ with the constraint that $\gamma>0$ for the (sub)-extensive law and $\gamma\leq0$ for the area law.
For each fit we also compute the sum of the squared residuals per degree of freedom ($E = \chi^2/\mathsf{dof}$), which quantifies how well the fit matches the data.
Hence for both entanglement and magic, all chosen values of $p$, and all chosen values of $\eta$ we have the quantities $E_\mathrm{ext}$ and $E_\mathrm{area}$ that quantify how well the data matches a (sub)--extensive and area law, respectively.
Finally, using a linear interpolation between the discrete set of the $p$ values, we quote the value of $p$ for which both fits work equally well as the critical measurement rate.

Figure~\ref{fig:phase-diagram} shows the result of this analysis for several different values of $\eta$.
The data suggest that the entanglement phase transition is unaffected by the presence of a vanishing density of T--gates, while the magic phase transition shifts depending on the prefactor $\eta$ of that vanishing density.
For the studied range of $\eta$ between $0.5$ and $8.0$ this results in phase I for low measurement rates, phase II for intermediate rates, and phase III for higher rates.
Figure~\ref{fig:phase-diagram} naturally prompts the question whether the transition rates stay apart for smaller values of $\eta>0$, merge, or even cross giving rise to phase IV.
This regime is, however, difficult to access numerically due to the very sparse occurrences of T--gates.
Sparser occurrences of T--gates in the circuit demands a larger number of trajectories and a longer simulation time to reach a steady state value.


\begin{figure}[t]
	\includegraphics[width=0.48\textwidth]{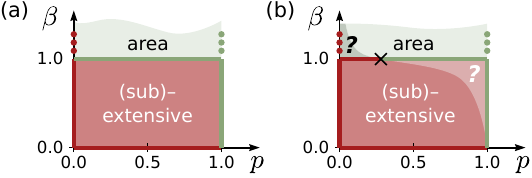}
	\caption{\label{fig:phase-sketches}%
		Phase diagram ($\beta$ vs. $p$) of magic for the separable hybrid quantum circuit~(a) and a sketch of the conjectured phase diagram for the full hybrid quantum circuit~(b).
		While preliminary numerical results for the full circuit indicate that most of the region below $\beta<1$ follows a (sub)--extensive law it is still unclear whether there exists a transition to an area scaling with $p_c^\mathrm{magic}<1$ in that region.
	}
\end{figure}%

\textit{Considerations for the $\beta \neq 1$ cases.}---%
A natural question following from the above result is whether the non--trivial phase transition we observe persists for a T--gate density that vanishes as $\eta/N^\beta$ even if $\beta \neq 1$.
As a first attempt to answer this question we consider a simple hybrid quantum circuit identical to the circuit considered above but with the two qubit brickwork structure replaced by only single site random Clifford gates.
In this scenario the total state is always a product state of single qubits.
This allows straight forward analytical considerations that (for $\beta > 0$) yield
\begin{equation} \label{eq:magic-non-entangling}
\bar{\mathcal{M}}_\mathrm{ss}(N \rightarrow \infty) = \eta \, \bar{\mathcal{M}}_1 \, \frac{1-p}{p} \, N^{1-\beta} \mathrm{,}
\end{equation}
where $\bar{\mathcal{M}}_1$ is the average magic of a random single site stabilizer state after the application of a single T--gate.
Hence, assuming a non-trivial measurement rate $0<p<1$, we find a sub--extensive law for \mbox{$0 < \beta < 1$, a $p$} dependent constant for $\beta=1$, and a vanishing magic for $1 < \beta$.

In Fig.~\ref{fig:phase-sketches} we sketch the phase diagram of magic as a function of the measurement rate $p$ and the scaling parameter $\beta$ for a fixed prefactor $\eta=1.0$.
Figure~\ref{fig:phase-sketches}a shows the phase diagram for the exactly solvable separable model, while Fig.~\ref{fig:phase-sketches}b shows a conjectured phase diagram for the full quantum circuit (including the two qubit Clifford brickwork structure), for which only the $\beta=1.0$ line is known through the computations and analysis carried out above.
Finally we note, that upon comparison of the non--entangling $\beta=1.0$ case with the phase transition of the brickwork circuit above it appears that the entangling gates effectively protect magic from destruction by measurements.


\textit{Conclusions.}---%
In this work we find a MIPT in magic that is independent of the known MIPT in entanglement.
A key ingredient for this result is the controlled injection of non--Clifford resources into the circuit, for which we have chosen Clifford circuits interspersed with projective measurements and T--gates.
We note that also other classes of hybrid quantum circuits with different forms of non--Clifford resources (such as general phase shifts or measurements under some angle relative to the computational basis) may show similarly interesting behavior.
We believe that our results and further investigation of the questions raised in this paper are beneficial to gain a deepened understanding of MIPTs in hybrid quantum circuits and their relation to computational quantum advantage and quantum error correction.


\textit{Acknowledgments.}---%
We thank Lorenzo Piroli, Xhek Turkeshi, and Christopher White for helpful comments on the manuscript.
Also, we thank Titas Chanda, Mario Collura, Piotr Sierant, Poetri Tarabunga, and Xhek Turkeshi, for many insightful discussions.
G.\,E.\,F. and R.\,F. acknowledge support from  ERC under grant agreement n.101053159 (RAVE).
M.\,D. and E.\,T. were partly supported by the MIUR Programme FARE (MEPH), by QUANTERA DYNAMITE PCI2022-132919, and by the EU-Flagship programme Pasquans2.
M.\,D. and R.\,F. were partly supported by the PNRR MUR project PE0000023-NQSTI.
M.\,D. and R.\,F. work was in part supported by the International Centre for Theoretical Sciences (ICTS) for participating in the program -  Periodically and quasi-periodically driven complex systems (code: ICTS/pdcs2023/6).


\textit{Note added:}
After uploading this manuscript to arXiv, we became aware of a related work by Bejan, McLauchlan and Beri~\cite{bejan_dynamical_2023} that also discusses the interplay of magic and entanglement transitions in random unitary circuits.
While the details of the works largely differ, they both observe that entanglement and magic transitions are in general distinct.


%


\clearpage
\section{Supplemental Material}

We present additional information on (1) the properties of the stabilizer entropy, (2) the tensor network computations, and (3) the extraction of the critical measurement rates, supplementing the main text.

\renewcommand{\theequation}{S\arabic{equation}}
\renewcommand{\thefigure}{S\arabic{figure}}
\setcounter{equation}{0}
\setcounter{figure}{0}

\newcommand{\mainref}[2]{#1}

\subsection{Properties of the stabilizer entropy}
The stabilizer R\'enyi entropy is a recently introduced non--stabilizerness monotone~\cite{leone_stabilizer_2022}.
It can be computed numerically even for a large number of qubits~\cite{haug_quantifying_2023, lami_quantum_2023, tarabunga_many-body_2023} and is also experimentally accessible~\cite{oliviero_measuring_2022}.
In this section, we briefly state some of its key properties to allow easy access to the main results of the paper. 

For three common choices of $\alpha$ the stabilizer R\'enyi entropy (as defined in Eq.~\eqref{eq:sre-def} of the main text) reads
\begin{equation}  
	\mathcal{M}_\alpha(|\psi\rangle) =
	\begin{cases}
		\log_2 \left( \lvert \left \lbrace  P \in \mathcal{P}_N :  \langle P \rangle_\psi \neq  0  \right \rbrace \lvert \right)-N & \alpha \rightarrow 0 \\
		-\sum_P 2^{-N} \langle P \rangle_\psi^2 \log_2 \left( \langle P \rangle_\psi^2 \right) & \alpha \rightarrow 1 \\
		-\log_2 \left(\sum_P 2^{-N} \langle P \rangle_\psi^4   \right) & \alpha =2
	\end{cases}
\end{equation}
where $P\in\mathcal{P}_N$ is the group of all $N$--qubit Pauli strings with +1 phases.
We list some key properties of the stabilizer $\alpha$--R\'eney entropies, alongside the references that contain the respective proofs:
\begin{enumerate}[label=(\roman*)]
    \item \emph{Faithfulness}: $\mathcal{M}_\alpha \left(|\psi\rangle \right)=0$ if and only if $|\psi \rangle$ is a stabilizer state (see Ref.~\cite{leone_stabilizer_2022}).
    \item \emph{Stability under free operations}: For any unitary Clifford operator $C$ and state $|\psi\rangle$ it holds that \mbox{$\mathcal{M}_\alpha(C |\psi \rangle) = \mathcal{M}_\alpha\left( |\psi \rangle \right)$} (see Ref.~\cite{leone_stabilizer_2022}).
    \item \emph{Additivity}: $\mathcal{M}_\alpha \left( |\psi \rangle \otimes |\phi \rangle \right) = \mathcal{M}_\alpha \left(|\psi \rangle \right)+ \mathcal{M}_\alpha \left(|\phi \rangle \right)$ (see Ref.~\cite{leone_stabilizer_2022}).
    \item \emph{Bounded}: For any $N$-qubit state $|\psi\rangle$ it holds that $0\leq \mathcal{M}_\alpha(|\psi\rangle)<N$ (see Ref.~\cite{leone_stabilizer_2022}).
    \item  $M_{\alpha^{\prime}}(|\psi\rangle)<\mathcal{M}_\alpha(|\psi\rangle)$ for $\alpha^{\prime}>\alpha$ (see Ref.~\cite{haug_stabilizer_2023}).
    \item The stabilizer entropies consitute a lower bound to the so--called $T$-count $t(|\psi\rangle)$ of a state: \mbox{$\mathcal{M}_\alpha\left( |\psi \rangle \right)< t(|\psi\rangle)$} (see Ref.~\cite{gu_little_2023}).
    \item For $\alpha > 1/2$ the stabilizer entropies constitute  a lower bound to the so--called robustness of magic: $\mathcal{M}_\alpha\left( |\psi \rangle \right)<\mathcal{R}_{\psi}$ (see Refs.~\cite{heinrich_robustness_2019, leone_stabilizer_2022}).
\end{enumerate}

\subsection{Additional information on the tensor network computations}
Figures~\ref{fig:entropies-vs-size}~and~\ref{fig:phase-diagram} in the main text are based on tensor network calculations of 512 trajectories for each combination of various T-gate scaling prefactors $\eta$, measurement rates $p$, and system sizes $N$.
The sets of possible values are  $\eta \in \{0.5, 1.0, 2.0, 4.0, 8.0\}$,  $p \in \{0.14, 0.16, \ldots, 0.34\}$, and $N \in \mathcal{N} = \{8, 12, 16, 24, 32, 48, 64, 92, 128, 184\}$, where we however exclude $N \in  \{92, 128, 184\}$ for $p \in \{0.14, 0.16\}$.
All computations assume $\beta=1$.
We start each trajectory in the $|00\ldots0\rangle$ state and evolve it as a matrix product state (MPS) using the Julia language version of the tensor network package iTensor~\cite{vidal_efficient_2004, fishman_itensor_2022}.
Each of the random 2--site Clifford gates are generated independently by 30 alternating layers of a random choice of (1) random single site Clifford gates, (2) the swap operation, (3) the controlled--not operation acting on the left qubit, and (4) the controlled--not operation acting on the right qubit.
The single site Clifford gates are directly sampled from the 24 different gates of the Clifford group.
As a singular value truncation strategy we use a combination of both a relative and an absolute threshold of $\epsilon=10^{-6}$ and $\chi=256$, respectively.
To check convergence with respect to these thresholds we have performed the same computations (for a subset of parameter sets) substituting the thresholds with $\epsilon=10^{-5}$ and $\chi=128$.
We found that the results did not change significantly compared to the errors due to the statistics of the finite number of trajectories.

\begin{figure}[t]
	\includegraphics[width=0.45\textwidth]{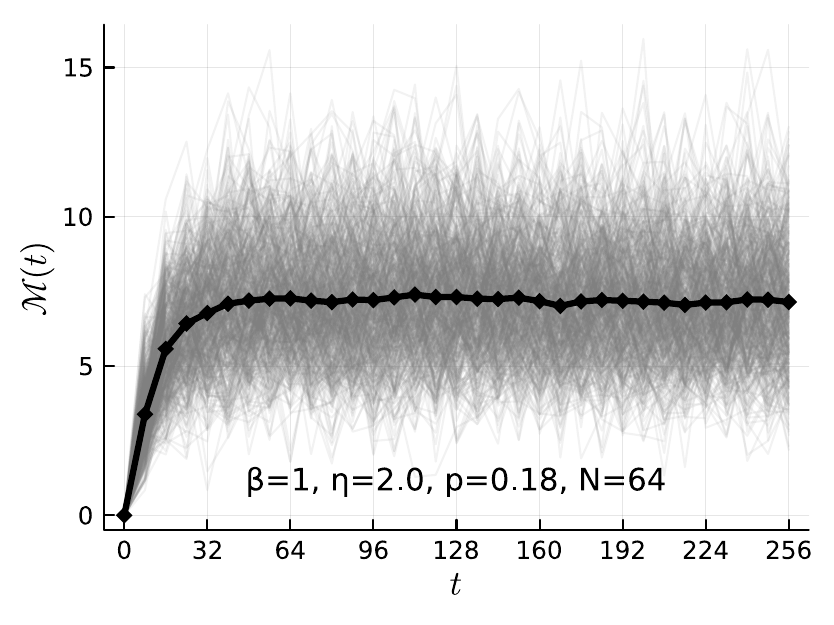}
	\caption{\label{fig:trajectories}%
		Time evolution of the stabilizer 2--R\'eny entropy $\mathcal{M}(t)$ of 512 trajectories (gray lines) and their average (thick solid black line).
	}
\end{figure}%

For every qubit and every time step we independently apply a T--gate with probability $q = \eta/N^\beta$.
The application of a measurement occurs analogously with probability $p$.
The outcome of each measurement is, however, weighted by the born rule and we renormalize the state after the projection.

At every $8^\mathrm{th}$ time step we compute the Schmidt coefficients between the left and the right half of the chain, as well as the Pauli string distribution.
The Schmidt coefficients are obtained by bringing the orthogonality center of the MPS to site $N/2$ and performing a singular value decomposition between site $N/2$ and $N/2+1$.
To obtain the Pauli string distribution we implement the method explained in~\cite{lami_quantum_2023}.
We found that a sample size of 128 Pauli strings is sufficient for a good estimate of the stabilizer 2--R\'eny entropy.
From the Schmidt coefficients and the Pauli string distribution we calculate the von-Neumann entropy $\mathcal{E}(t)$ and the stabilizer 2--R\'eny entropy $\mathcal{M}(t)$.

\begin{figure}[t]
	\includegraphics[width=0.45\textwidth]{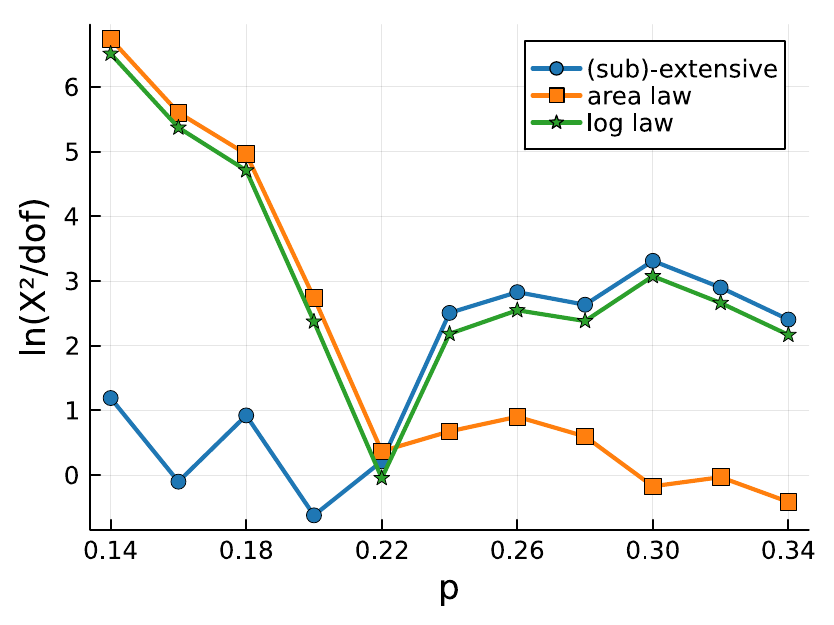}
	\caption{\label{fig:-scaling-fits}%
		The squared residuals per degree of freedom as a function of the measurement rate $p$ for the magic of a hybrid circuit with T--gate density $q=\eta/N^\beta$, where $\eta=2.0$ and $\beta=1.0$.
	}
\end{figure}%

In Fig.~\ref{fig:trajectories} we show the 512 trajectories of $\mathcal{M}(t)$ as gray lines and their average $\bar{\mathcal{M}}(t)$ for $\eta=2.0$, $p=0.18$, and $N=64$.
We can see that $\bar{\mathcal{M}}(t)$ reaches a steady state after about 50~time steps for these parameters.
For all parameter sets considered in this paper we find that the approximate steady state is reached before $t=2N$ time steps.
To extract the approximate steady state value of $\bar{\mathcal{M}}(t)$ and $\bar{\mathcal{E}}(t)$ we average over all values of $\mathcal{M}(t)$ and $\mathcal{E}(t)$ from time $t=2N$ to $t=4N$.
Because we have each one value of $\bar{\mathcal{M}}(t)$ and $\bar{\mathcal{E}}(t)$ for every $8^\mathrm{th}$ time step this means that the approximate steady states $\bar{\mathcal{M}}_\mathrm{ss}$ and $\bar{\mathcal{E}}_\mathrm{ss}$ are determined as an average over $P=N/4$ values.
The error bar for each of these approximate steady states shown in Fig.~\ref{fig:entropies-vs-size} of the main text is the standard deviation of this distribution.
The computations were performed on the Cineca HPC Marconi cluster with 2$\times$24-cores Intel Xeon 8160 CPUs and consumed a total of 36k CPUh.

\subsection{Additional information on the extraction of critical measurement rates}

\begin{figure}[t]
	\includegraphics[width=0.45\textwidth]{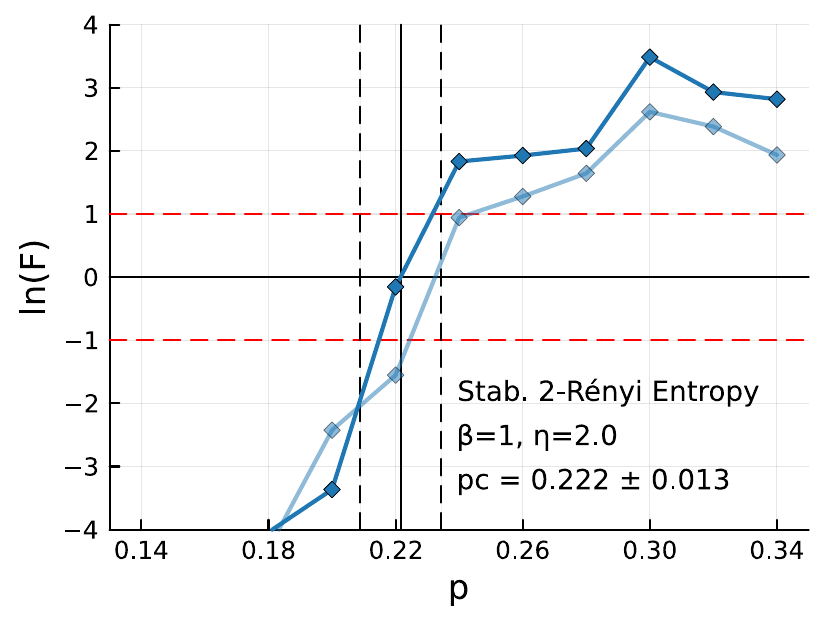}
	\caption{\label{fig:-scaling-F-test}%
		The logarithm of $F$ as a function of the measurement rate $p$ for the magic of a hybrid circuit with T--gate density $q=\eta/N^\beta$, where $\eta=2.0$ and $\beta=1.0$.
		The blue diamond symbols show $\ln[F(p)]$ calculated for the discrete set of computed measurement rates $p$ and the solid blue line shows a linear interpolation. The faint blue diamonds and faint solid blue line show the same calculated from the same data, except that the data points corresponding to the largest system size have been excluded.
		The vertical solid black line shows the extracted critical measurement rate $p_c$.
		The vertical dashed black lines indicates the estimated error $\sigma_{p_c}$.
	}
\end{figure}%

As explained in the main text we fit the data computed in the previous section to the function $f(N) = a + b N^\gamma$ with the constraint that $\gamma>0$ for the (sub)-extensive law and $\gamma\leq0$ for the area law, to see which law fits the data best.
For each curve and fitting function we perform a weighted least square fit and extract the squared residuals per degree of freedom $\chi^2/\mathsf{dof}$, where
\begin{equation}
	\chi^2 = \sum_{N\in\mathcal{N}} \frac{\bar{\mathcal{M}}_\mathrm{ss}(N) - f(N,a_\mathrm{fit},b_\mathrm{fit},\gamma_\mathrm{fit})}{\sigma(N)},
\end{equation}
with the computed steady state expectation value of magic $\bar{\mathcal{M}}_\mathrm{ss}(N)$ and the error $\sigma(N)$ of each data point.
The degree of freedom ($\mathsf{dof}$) is the number of data points minus the number of fit parameters.
We find that using the standard deviation computed above often leads to a $\chi^2/\mathsf{dof} \simeq 0.05$ which signals that we have overestimated the error.
Assuming that the distribution of averaged values from time $t=2N$ to $t=4N$ is approximately uncorrelated, the standard deviation needs to be rescaled by a factor of $1/\sqrt{P}$.
In Fig.~\ref{fig:-scaling-fits} we plot $\chi^2/\mathsf{dof}$ against the measurement rate $p$ for the (sub)--extensive and area law scaling fit functions using the rescaled standard deviation.
In addition to this, we also plot $\chi^2/\mathsf{dof}$ for fits of a logarithmic scaling function $f_\mathrm{log}(N) = a + b \ln(N)$.
For each fit we use the data points corresponding to the 7 largest available system sizes.
We can clearly see that for small measurement rates the (sub)--extensive scaling fits the data best, while at large rates the area scaling fits best.
Moreover, using the rescaled standard deviation, for all $p$ the best fits have a $\chi^2/\mathsf{dof}$ close to the optimal value of 1.
To extract the critical measurement rate we plot in Fig.~\ref{fig:-scaling-F-test} the logarithm of
\begin{equation}
	F(p) = \frac{E_\mathrm{ext}(p)}{E_\mathrm{area}(p)}\mathrm{,}
\end{equation}
which is the ratio of $E(p)=\chi^2/\mathsf{dof}$ for the (sub)-extensive and area law scaling fits.
We connect the discrete data points with a linear interpolation and quote the measurement rate at which $\ln[F(p_c)]=0$ as the critical measurement rate $p_c$.
We estimate the error of this value $\sigma_{p_c}$ to be composed of two contributions with $\sigma_{p_c} = \sqrt{\sigma_A^2 + \sigma_B^2 }$.
The first contribution $\sigma_A$ is given by $p^\pm$ compared to $p_c$ for which $\ln[F(p^\pm)]=\pm 1$, i.e. $\sigma_A = \max\{|p-p_c| \text{~with~} \ln[F(p)]=\pm 1\}$.
The second contribution $\sigma_B$ is the absolute value of the difference between $p_c$ and the critical value $p^\prime_c$ computed from a restricted data set.
We do this by excluding the data points corresponding to the largest available system size.
We show the result of the analysis for this restricted data set as the fainted line in Fig.~\ref{fig:-scaling-F-test}.
The same analysis is done independently for both magic and entanglement, and for different T--gate density prefactors $\eta$.
Figure~\ref{fig:phase-diagram} in the main text summarizes the results of these calculations.


\end{document}